\documentclass[pra,onecolumn]{revtex4}
\usepackage{indentfirst}
\usepackage{graphicx}
\usepackage{amsmath}
\usepackage{amsfonts}
\usepackage{color}
\begin{document}
\title{Dynamics of Entanglement and Bell-nonlocality for Two Stochastic Qubits with Dipole-Dipole Interaction}
\author{Ferdi Altintas}
\author{Resul Eryigit}\email[email: ]{resul@ibu.edu.tr}
\affiliation{Department of Physics, Abant Izzet Baysal University, Bolu, 14280-Turkey.}
\begin{abstract}
We have studied the analytical dynamics of Bell nonlocality as measured by CHSH inequality and entanglement as measured by concurrence for two noisy qubits that have dipole-dipole interaction. The nonlocal entanglement  created by the dipole-dipole interaction is found to be protected from sudden death for certain initial states. 
\end{abstract}

\maketitle
\section{Introduction}
Entanglement is a kind of quantum correlation and refers to separability of the states. The entangled states are the main resource for quantum information and computation applications, such as  quantum teleportation~\cite{bennet1}, superdense coding~\cite{bennet2}, and quantum cryptography~\cite{ekert}. Entanglement and other types of quantum correlations are not easy to maintain when the system is in contact with an environment which leads to decoherence~\cite{decohere}. Thus the study of entangled states under decoherence is one the most important aspects in theoretical as well as experimental areas~\cite{wall,ameida}. It was shown that the interaction between the single qubit and its environment leads to exponential decay  of the qubit coherences while entanglement between two such qubits can cease to exist  in a finite time, a phenomenon which was named "entanglement sudden death" (ESD) by Yu and Eberly~\cite{esd}. Later Lopez {\it et al.,} revealed that when the bipartite entanglement suddenly disappears, the entanglement of the corresponding reservoir suddenly and necessarily appears, which is a counter phenomenon to ESD  and termed "entanglement sudden birth" (ESB)~\cite{esb}. For bipartite systems, many theoretical efforts have been devoted to ESD~\cite{esd,noiseonly,fare,xmatrix,cxp,aar,esdd} as well as ESB~\cite{esb,esb2,esb3} and also ESD is verified experimentally~\cite{ameida}. However, they need further clarifications for a deeper understanding.

On the other hand, it was demonstrated by Werner that there are some entangled mixed states whose correlations can be reproduced by a classical local model~\cite{werner}. The only way of identification of such states is to use violations of Bell-inequalities. The violations of Bell inequalities identify the genuine multipartite entanglement which is useful for quantum computations~\cite{acin1,acin2}. For a bipartite system, there are many Bell type inequalities, such as quadratic- and CHSH-Bell inequalities~\cite{uffink,chshinequ}. The CHSH-Bell inequality is a good indicator of nonlocal correlations and the relation with the entanglement has been already known~\cite{bfc,bfc2,ll,kk,miron,lx,mhk,mbfc}. Among them, Bellomo {\it et al.}, found that Bell inequality might not be violated for a state with high entanglement for a two qubit system subject to amplitude damping~\cite{bfc}. Kofman {\it et al.,} showed that the survival time for entanglement should be much longer than the Bell inequality violation under dephasing and energy relaxation type channels at an arbitrary temperature~\cite{kk}. And recently, Li {\it et al.}, studied the violations of Bell inequality under classical Ornstein-Uhlenbeck noise and showed that strong non-Markovian effect can protect the nonlocal entanglement identified by the violation of the CHSH inequality~\cite{ll}.

In this paper, we have analyzed the dynamics of Bell nonlocality as measured by CHSH inequality and entanglement as measured by concurrence for a system of two coupled qubits that interact via dipole-dipole interaction. The effect of environment on the qubit is modelled as a stochastic energy level with Ornstein-Uhlenbeck type correlation~\cite{noiseonly,fare}. The effects of non-Markovianity of the dynamics, purity of the initial states as well as the dipole-dipole interaction strength on Bell nonlocality and entanglement have been investigated for a system initially prepared in extended Werner-like states.

The organization of this paper is as follows. In Sec.~\ref{model}, we introduce the model and its solution by solving the master equation for the two qubit reduced density matrix. In Sec.~\ref{entbell}, we briefly discuss the Wootters concurrence as well as CHSH Bell inequality and the effect of purity, dipole-dipole interaction strength and the non-Markovianity on entanglement and Bell nonlocality is studied for extended Werner-like initial states. In Sec.~\ref{conc}, we conclude as a summary of important results.  

\section{The model and its solution}
\label{model}

In the present paper, we consider two interacting qubits that are coupled to their independent environments which leads to fluctuating energy levels. The qubit-qubit interaction is assumed to be of the type Heisenberg XX model.  This model can be thought of as the Kubo-Anderson model extended to two coupled qubits~\cite{kubo}. The typical Hamiltonian for this model can be given as~\cite{noiseonly,fare,dipoleint} (we set $\hbar=1$):
\begin{eqnarray}
\label{hamiltonian}
\hat{H}_{tot}(t)=J(\hat{\sigma}_x^A\hat{\sigma}_x^B+\hat{\sigma}_y^A\hat{\sigma}_y^B)+\frac{\Omega_A(t)}{2} \hat{\sigma}_z^A+\frac{\Omega_B(t)}{2} \hat{\sigma}_z^B,
\end{eqnarray}
where $J$ is the qubit-qubit interaction strength, $\hat{\sigma}_{i}^{A,B} (i=x,y,z)$ are the usual Pauli spin operators and $\Omega_{A,B}(t)$ are the independent frequency fluctuations of the qubits  with mean value properties that obey non-Markovian approximation:
\begin{eqnarray}
\label{mean}
M\{ \Omega_i(t)\}=0,\\
M\{\Omega_i(t)\Omega_i(s)\}&=&\alpha(t-s)\nonumber\\
&=&\frac{\Gamma_i \gamma}{2} e^{-\gamma \left|t-s\right|} ,\quad i=A,B,
\end{eqnarray}
where $M\{...\}$ stands for the statistical mean over the noise $\Omega_A(t)$ and $\Omega_B(t)$. Here $\Gamma_i (i=A,B)$ are the damping rates due to the coupling to the environments, $\gamma$ is the noise bandwidth which determines the environment's finite correlation time $(\tau_c=\gamma^{-1})$ and $\alpha(t-s)$ is the reservoir correlation function. For simplicity, we will take the noise properties to be the same for $A$ and $B$ (e.g., $\Gamma_A=\Gamma_B\equiv\Gamma$). And note that in the limit $\gamma\rightarrow\infty~(\tau_c\rightarrow0)$, Ornstein-Uhlenbeck noise reduces to the well-known Markovian case~\cite{xmatrix}:
\begin{eqnarray}
\label{markov}
\alpha(t-s)=\Gamma\delta(t-s).
\end{eqnarray}
For the total system described by the Hamiltonian~(\ref{hamiltonian}), the stochastic Schr\"{o}dinger equation is given by 
\begin{eqnarray}
\label{schrödinger}
i{d \over dt}\left|\Psi(t)\right\rangle=\hat{H}_{tot}(t)\left|\Psi(t)\right\rangle,
\end{eqnarray}
with formal solution 
\begin{eqnarray}
\label{solution}
\left|\Psi(t)\right\rangle=\hat{U}(t,\Omega_A,\Omega_B)\left|\Psi(0)\right\rangle,
\end{eqnarray}
where the stochastic propagator $\hat{U}(t,\Omega_A,\Omega_B)$ is given by
\begin{eqnarray}
\label{propagator}
\hat{U}(t,\Omega_A,\Omega_B)=e^{-i\int_0^t \hat{H}_{tot}(s)ds}.
\end{eqnarray}
The reduced density matrix for qubits $A$ and $B$ is then obtained from the statistical mean
\begin{eqnarray}
\label{reduceddensity}
\hat{\rho}=M\{ \left|\Psi(t)\right\rangle\left\langle \Psi(t)\right|\}.
\end{eqnarray}
With the help of the raising and lowering operators, $\hat{\sigma}_{\pm}^{A,B}=(\hat{\sigma}_x^{A,B}\pm i \hat{\sigma}_y^{A,B})/2$, and the stochastic Schr\"{o}dinger equation~(\ref{schrödinger}), the master equation for the reduced density matrix can be derived as~\cite{noiseonly,master1,master2,master3}:
\begin{eqnarray}
\label{mastereqn}
{d \over dt}\hat{\rho}=-i[\hat{H},\hat{\rho}]-\frac{G(t)}{2}(2 \hat{\rho}-\hat{\sigma}_z^A\hat{\rho}\hat{\sigma}_z^A-\hat{\sigma}_z^B\hat{\rho}\hat{\sigma}_z^B),
\end{eqnarray}
where $G(t)=\int_0^t\alpha(t-s)ds= \frac{\Gamma}{2}(1-e^{-\gamma t})$ is a time-dependent coefficient which includes the memory information of the environmental noise and $\hat{H}=2J (\hat{\sigma}_{+}^A\hat{\sigma}_{-}^B+\hat{\sigma}_{-}^A\hat{\sigma}_{+}^B)$ is the interaction Hamiltonian which represents the dipole-dipole interaction between the qubit $A$ and $B$~\cite{dipole}. Note that the first term on the right-hand side of Eq.~(\ref{mastereqn}) leads to oscillatory dynamics while the second term causes decay.

The differential equations governing the time evolution of the system in the standard basis $\{\left|1\right\rangle\equiv\left|11\right\rangle,\left|2\right\rangle\equiv\left|10\right\rangle,\left|3\right\rangle\equiv\left|01\right\rangle,\left|4\right\rangle\equiv\left|00\right\rangle\}$ can be easily calculated:
\begin{eqnarray}
\label{maselements}
\dot{\rho}_{ii}&=&0\quad(i=1,4),\nonumber\\
\dot{\rho}_{22}&=&2iJ(\rho_{23}-\rho_{23}^*),\nonumber\\
\dot{\rho}_{33}&=&-2iJ(\rho_{23}-\rho_{23}^*),\nonumber\\
\dot{\rho}_{12}&=&2iJ\rho_{13}-G(t)\rho_{12},\nonumber\\
\dot{\rho}_{13}&=&2iJ\rho_{12}-G(t)\rho_{13},\nonumber\\
\dot{\rho}_{14}&=&-2G(t)\rho_{14},\nonumber\\
\dot{\rho}_{23}&=&2iJ(\rho_{22}-\rho_{33})-2G(t)\rho_{23},\nonumber\\
\dot{\rho}_{24}&=&-2iJ\rho_{34}-G(t)\rho_{24},\nonumber\\
\dot{\rho}_{34}&=&-2iJ\rho_{24}-G(t)\rho_{34},
\end{eqnarray}
where the asterisk in the superscript of $\rho_{mn}^*$ denotes the complex conjugation of $\rho_{mn}$. After a simple calculation, the analytical solutions of the coupled first order ordinary differential equations in Eq.~(\ref{maselements}) can be found. They are reported in~\ref{App1}. It should be noted that in the limit of $J\rightarrow0$, the solutions~(\ref{solmaselements}) have simple analytical forms which are analyzed in Refs.~\cite{noiseonly} and~\cite{ll}.

\section{Entanglement and CHSH-Bell Inequality Dynamics}
\label{entbell}

For two-qubit systems, Wootters concurrence can be used as a measure of entanglement~\cite{wootters}. The concurrence function varies from $C=0$ for a separable state to $C=1$ for a maximally entangled state. The concurrence function may be calculated from the density matrix $\hat{\rho}$ for qubits $A$ and $B$ as:
\begin{eqnarray}
\label{concurrence}
C(\hat{\rho})=\max\{0,\sqrt{\lambda_1}-\sqrt{\lambda_2}-\sqrt{\lambda_3}-\sqrt{\lambda_4}\},
\end{eqnarray}
where the quantities $\lambda_i$ are the eigenvalues in decreasing order of the matrix
\begin{eqnarray}
\label{rhotrans}
\hat{\rho}_{trans}=\hat{\rho}(\hat{\sigma}_y^A\otimes\hat{\sigma}_y^B)\hat{\rho}^*(\hat{\sigma}_y^A\otimes\hat{\sigma}_y^B).
\end{eqnarray}

In the following, we consider entanglement dynamics of the qubits whose density matrix has a common X-form~\cite{xmatrix}:
\begin{eqnarray}
\label{xmatrix}
\hat{\rho}=\left [ \begin{array}{cccc}
  \rho_{11}  & 0 & 0  & \rho_{14} \\
  0  & \rho_{22} & \rho_{23}  & 0\\ 0  & \rho_{32} & \rho_{33}  & 0 \\ \rho_{41}  & 0 & 0  & \rho_{44}
\end{array} \right] \ .
\end{eqnarray}
Such X-states arise in a wide variety of physical situations and include pure Bell states~\cite{xmatrixuse} as well as the well-known Werner mixed states~\cite{werner}. Note that the time evolution of Eq.~(\ref{maselements}) preserves the form of X-state. Then one can easily show that the concurrence function for the X-state~(\ref{xmatrix}) is given by~\cite{noiseonly}
\begin{eqnarray}
\label{xconcurrence}
C(\hat{\rho})=2\max\{0,\left|\rho_{23}\right|-\sqrt{\rho_{11}\rho_{44}}, \left|\rho_{14}\right|-\sqrt{\rho_{22}\rho_{33}}\}.
\end{eqnarray}

For two-qubit case the CHSH-type Bell inequality can be written in the following form~\cite{chshinequ}
\begin{eqnarray}
\label{bell1}
B(\hat{\rho})=\left|\left\langle O_AO_B\right\rangle-\left\langle O_AO'_B\right\rangle\right|+\left\langle O'_AO_B\right\rangle+\left\langle O'_AO'_B\right\rangle\leq 2,
\end{eqnarray}
where $\left\langle O_AO_B\right\rangle=Tr(\hat{\rho}O_AO_B)$ is the correlation function and $O_S={\bf O_S}.{\bf \sigma_S}$, wherein ${\bf O_S}=(\sin\theta_S\cos\phi_S,\sin\theta_S\sin\phi_S,\cos\theta_S)$ is the unit vector, ${\bf \sigma_S}=(\sigma_1^S,\sigma_2^S,\sigma_3^S)$ is the Pauli matrices vector and $O'_S=O_S(\theta'_S,\phi_S)$. Bellomo {\it et al.}, showed that the maximum of the Bell function $B(\hat{\rho})$ for a X-structured density matrix~(\ref{xmatrix}) can be given as~\cite{bfc,bfc2}:
\begin{eqnarray}
\label{bell2}
B_{max}(\hat{\rho})=2\sqrt{P^2(t)+Q^2(t)},
\end{eqnarray}
where 
\begin{eqnarray}
\label{pq}
P(t)&=&\rho_{11}+\rho_{44}-\rho_{22}-\rho_{33},\nonumber\\
Q(t)&=&2(|\rho_{14}|+|\rho_{23}|).
\end{eqnarray}
It should be noted that in the regions where $B_{\max}(\hat{\rho})>2$, the CHSH inequality is violated. It implies that the state $\hat{\rho}$ is genuinely bipartite Bell nonlocal, thus the correlations cannot be accessible by any classical local model. Also notice that the expression for $B_{\max}(\hat{\rho})$ in Eq.~(\ref{bell2}) coincides with the one that would be obtained using the formal Horodecki criterion~\cite{hhh}.

In the following, we consider two classes of initial states, called extended Werner-like (EWL) states in the form:
\begin{eqnarray}
\label{ewl}
\hat{\rho}_{\Phi}(0)&=&\frac{1-r}{4}I_4+r\left|\Phi\right\rangle\left\langle \Phi\right|\nonumber\\
&=&\left(
\begin{array}{cccc} \frac{1-r}{4} & 0 & 0 & 0  \\ 0 & \frac{1-r}{4}+\alpha^2r & \alpha\beta r & 0  \\ 0 & \alpha\beta r & \frac{1-r}{4}+\beta^2r & 0  \\  0  & 0 & 0 & \frac{1-r}{4}\end{array} \right),\nonumber\\
\hat{\rho}_{\Psi}(0)&=&\frac{1-r}{4}I_4+r\left|\Psi\right\rangle\left\langle \Psi\right|\nonumber\\
&=&\left(
\begin{array}{cccc} \frac{1-r}{4}+\alpha^2r & 0 & 0 & \alpha\beta r \\ 0 & \frac{1-r}{4} & 0 & 0  \\ 0 & 0 & \frac{1-r}{4}& 0  \\  \alpha\beta r  & 0 & 0 & \frac{1-r}{4}+\beta^2r \end{array} \right),
\end{eqnarray}
where $r$ is the purity of the initial states which ranges from 0 for maximally mixed states to 1 for pure states, $I_4$ is the $4\times4$ identity matrix, $\left|\Phi\right\rangle=\alpha\left|10\right\rangle+\beta\left|01\right\rangle, \left|\Psi\right\rangle=\alpha\left|11\right\rangle+\beta\left|00\right\rangle$ are the Bell-like pure states and the parameter $\alpha$ is sometimes called the degree of entanglement~(here we set $\alpha$ as real number and $\beta=\sqrt{1-\alpha^2}$)~\cite{bfc}. From Eq.~(\ref{maselements}) or~(\ref{solmaselements}), it can be easily noted that the initial states of Eq.~(\ref{ewl}) belong to "X" states at any time t. So the time-dependent concurrence and the maximum of the Bell function can be obtained from the the Eqs.~(\ref{xconcurrence}) and~(\ref{bell2}), respectively.

Before starting our qualitative analysis, we want to emphasize that the time-dependent concurrence and the maximum of the Bell function for $\hat{\rho}_{\Psi}(0)$ have simple analytical forms. According to Eq.~(\ref{solmaselements}), for this initial state the density matrix at any time can be found as:
\begin{eqnarray}
\label{dmepsi} 
\hat{\rho}_{\Psi}(t)&=&\left(
\begin{array}{cccc} \frac{1-r}{4}+\alpha^2r & 0 & 0 & \alpha\beta re^{-2f(t)}  \\ 0 & \frac{1-r}{4} & 0 & 0  \\ 0 & 0 & \frac{1-r}{4}& 0  \\  \alpha\beta re^{-2f(t)}  & 0 & 0 & \frac{1-r}{4}+\beta^2r \end{array} \right),
\end{eqnarray}
then the time-dependent concurrence and the maximum of Bell function for the $\hat{\rho}_{\Psi}(0)$ initial state can be obtained as:
\begin{eqnarray}
\label{cbpsi}
C^{\Psi}(\hat{\rho})&=&2\max\{0,r\alpha\beta e^{-2f(t)}-\frac{1-r}{4}\},\nonumber\\
B^{\Psi}_{\max}(\hat{\rho})&=&2r\sqrt{1+4(\alpha\beta e^{-2f(t)})^2},
\end{eqnarray}
where $f(t)=\frac{\Gamma}{2}\left(t+\frac{1}{\gamma}(e^{-\gamma t}-1)\right)$. For the initial state $\hat{\rho}_{\Phi}(0)$, a similar expression to Eq.~(\ref{cbpsi}) can be written by using the solution in~\ref{App1}. But the elements are quite involved and we do not display them here for brevity; $\hat{\rho}_{\Phi}(t)$ for special values of $\alpha$ and $J$ are displayed and discussed below. One should note that the dynamics of $\hat{\rho}_{\Psi}(0)$ initial state is undisturbed by the dipole-dipole interaction; which is expected because of the single excitation nature of dipole-dipole interaction and the form of $\left|\Psi\right\rangle$ state which is $\left|\Psi\right\rangle=\alpha\left|11\right\rangle+\sqrt{1-\alpha^2}\left|00\right\rangle$. Furthermore, Eq.~(\ref{cbpsi}) also corresponds to the time-dependent analytical expressions of $C(\hat{\rho})$ and $B_{\max}(\hat{\rho})$ for the initial states $\hat{\rho}_{\Phi}(0)$ and $\hat{\rho}_{\Psi}(0)$ in the absence of dipole-dipole interaction~(i.e., $J=0$)~\cite{ll}. The dynamics of entanglement and Bell nonlocality for the initial state, $\hat{\rho}_{\Psi}(0)$, was extensively examined in Ref.~\cite{ll}, thus we will not cover this state in detail at the rest of this paper. 
\begin{figure}[!ht]\centering
\includegraphics[width=6.5cm]{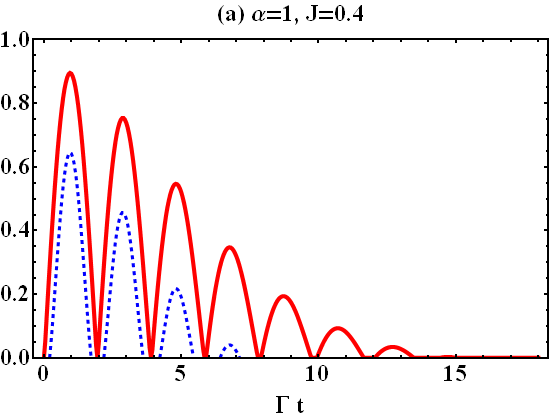}
\includegraphics[width=6.5cm]{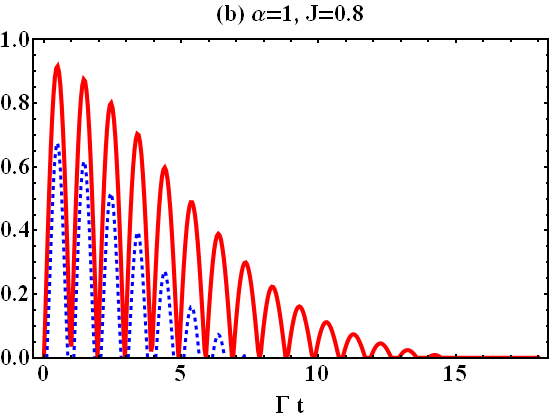}

\includegraphics[width=6.5cm]{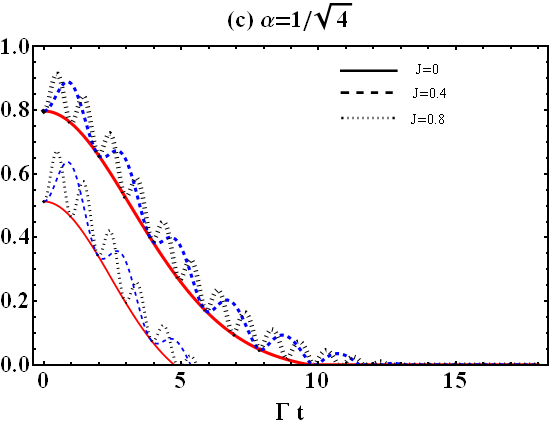}
\caption{\label{fig.1} Fig.~1(a) and~1(b) show $C(\hat{\rho})$~(solid plot) and $B_{\max}(\hat{\rho})-2$~(dashed plot) versus $\Gamma t$ for $\alpha=1$ and $J=0.4$~(Fig.~(a)) and $J=0.8$~(Fig.~(b)). Fig.~1(c) shows $C(\hat{\rho})$~(thick plot) and $B_{\max}(\hat{\rho})-2$~(thin plot) versus $\Gamma t$ for $\alpha=1/\sqrt{4}$ and $J=0$~(solid plot), $J=0.4$~(dashed plot) and $J=0.8$~(dotted plot). Here the figures are plotted for $\hat{\rho}_{\Phi}(0)$ with $r=0.95$ and $\gamma/\Gamma=0.1$.}
\end{figure}

The dynamics of concurrence and Bell nonlocality as a function of the dipole-dipole interaction strength and the dimensionless time are displayed in Fig.~1(a) and~(b) for $\alpha=1$ and in Fig.~1(c) for $\alpha=1/\sqrt{4}$ for the initial states $\hat{\rho}_{\Phi}(0)$ with $r=0.95$. All figures and quantities show a faster than exponentially decaying oscillatory behaviour with $J$-dependent oscillation frequency. Based on Eq.~(\ref{cbpsi}), for $\alpha=1$, it should be noted that in the absence of dipole-dipole interaction~(i.e., $J=0$), the states $\hat{\rho}_{\Psi}(t)$ and $\hat{\rho}_{\Phi}(t)$ have no entanglement and  do not violate CHSH inequality at any time, while in the presence of dipole-dipole interaction~(i.e., $J\neq0$), $\hat{\rho}_{\Phi}(t)$ possesses a high degree of entanglement and violates the CHSH-Bell inequality as can be seen from Fig.~1(a) and~(b). On the other hand, for $\alpha=1/\sqrt{4}$, the initial state $\hat{\rho}_{\Phi}(0)$ is entangled at $t=0$ and the entanglement as well as Bell nonlocality for $J\neq0$ live longer compared to the  $J=0$ case~(see Fig.~1(c)). The common feature of all figures is that Bell nonlocality is found to die out before the entanglement does as also found by Kofman and Korotkov in the case of dephasing and dissipative dynamics of two qubits~\cite{kk}.
\begin{figure}[!ht]\centering
\includegraphics[width=6.5cm]{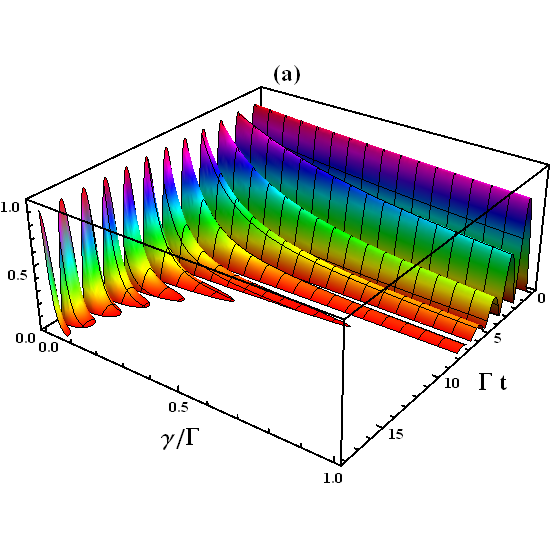}
\includegraphics[width=6.5cm]{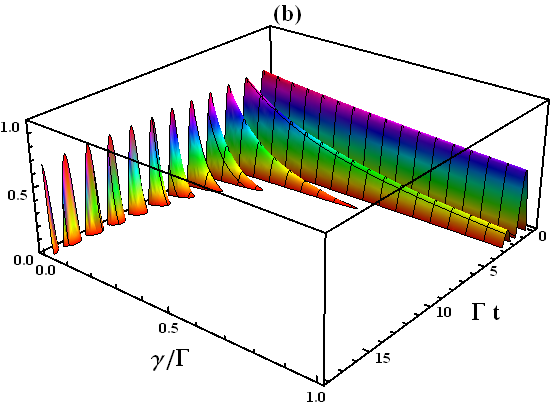}
\caption{\label{fig.2} $C(\hat{\rho})$~(Fig.~(a)) and $B_{\max}(\hat{\rho})-2$~(Fig.~(b)) versus $\Gamma t$ and $\gamma/\Gamma$ for $\hat{\rho}_{\Phi}(0)$ with $J=0.5, \alpha=1$ and $r=0.95$.}
\end{figure}

Non-Markovianity dependence of the entanglement and Bell inequality violation for $\hat{\rho}_{\Phi}(0)$ with $\alpha=1$ is displayed in Fig.~2(a) and~2(b), respectively. For the relatively high purity~($r=0.95$) considered in this case, the effect of non-Markovianity is found to be a prolonging of the lifetime of both concurrence and the Bell inequality violation. This finding is in line with many similar studies on the effect of non-Markovianity on the dynamics of entanglement, since non-Markovianity is a measure of memory effects, its increase may lead to a longer lived or protected entanglement~\cite{noiseonly,fare,esdd,bfc,bfc2,ll}.

\begin{figure}[!ht]\centering
\includegraphics[width=6.5cm]{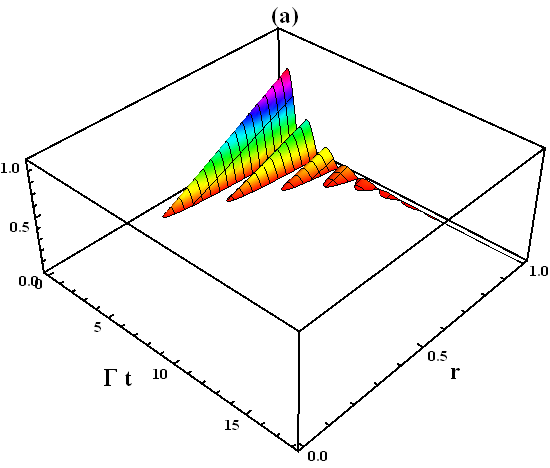}
\includegraphics[width=6.5cm]{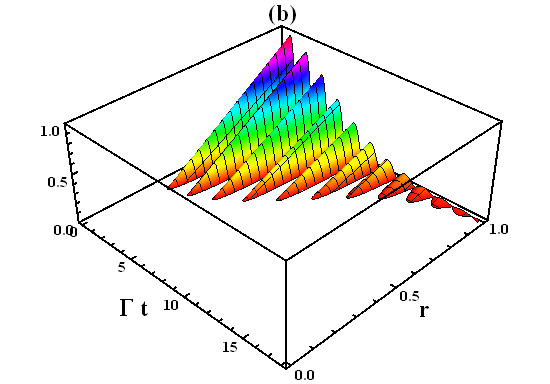}

\includegraphics[width=6.5cm]{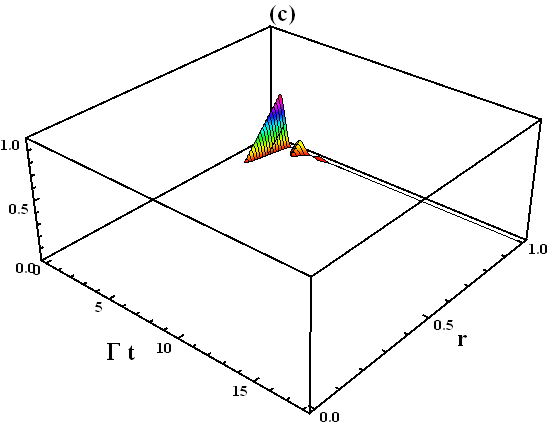}
\includegraphics[width=6.5cm]{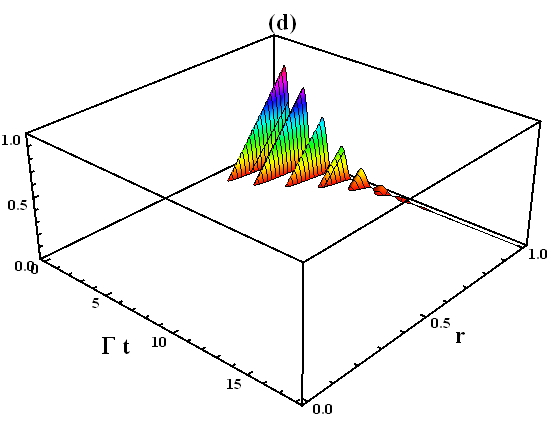}
\caption{\label{fig.3} $C(\hat{\rho})$~(Fig.~(a) and~(b)) and $B_{\max}(\hat{\rho})-2$~(Fig.~(c) and~(d)) versus $\Gamma t$ and $r$ for $\hat{\rho}_{\Phi}(0)$ with $J=0.5, \alpha=1$. Here Fig.~(a) and~(c) correspond to Markovian dynamics with $\gamma/\Gamma=10$ and Fig.~(b) and~(d) to non-Markovian dynamics with $\gamma/\Gamma=0.1$.}
\end{figure}

The effect of purity of the initial state on the dynamics of $\hat{\rho}_{\Phi}(t)$ state for Markovian and non-Markovian environments is displayed  in Figs.~3(a)-(d) for the concurrence and Bell inequality with $\alpha=1$. It should be noted that the dipole-dipole interaction creates a significantly wide purity range for non-zero entanglement  as $0.4<r\leq1$ for Markovian and $0.35<r\leq1$ for non-Markovian dynamics. On the other hand, the purity dependence of non-zero Bell nonlocality is narrower compared to the entanglement as $0.8<r\leq1$ and $0.7<r\leq1$ for Markovian and non-Markovian dynamics, respectively. The figures also show a pronounced difference between the dynamical behaviour of the both quantities for the pure and the mixed states. While both entanglement and Bell nonlocality decreases exponentially with time for pure states under Markovian and non-Markovian dynamics, there is a sudden death phenomenon for both quantities for the initially mixed states.

For the special case of $\alpha=1/\sqrt{2}$, the initial states considered here become
\begin{eqnarray}
\label{purebell}
\hat{\rho}_{\Phi}(0)&=&\frac{1-r}{4}I_4+r\left|\Phi\right\rangle\left\langle \Phi\right|,\nonumber\\
\hat{\rho}_{\Psi}(0)&=&\frac{1-r}{4}I_4+r\left|\Psi\right\rangle\left\langle \Psi\right|,
\end{eqnarray}
where $\left|\Phi\right\rangle=\frac{1}{\sqrt{2}}(\left|10\right\rangle+\left|01\right\rangle)$ and $\left|\Psi\right\rangle=\frac{1}{\sqrt{2}}(\left|11\right\rangle+\left|00\right\rangle)$ are the pure Bell states. With the help of the solutions presented in~\ref{App1}, the time-dependent density matrix for $\hat{\rho}_{\Phi}(0)$ can be written as:
\begin{eqnarray}
\label{tdps}
\hat{\rho}_{\Phi}(t)=\left(
\begin{array}{cccc} \frac{1-r}{4} & 0 & 0 & 0  \\ 0 & \frac{1+r}{4} & \frac{r}{2}e^{-2f(t)}  & 0  \\ 0 & \frac{r}{2}e^{-2f(t)}  & \frac{1+r}{4} & 0  \\  0& 0 & 0 & \frac{1-r}{4}  \end{array} \right),
\end{eqnarray}
then the concurrence and the maximum of the Bell function for the $\hat{\rho}_{\Phi}(t)$ of Eq.~(\ref{tdps}) can be expressed in a simple analytic form as  \begin{eqnarray}
\label{cbphsi}
C^{\Phi}(\hat{\rho})&=&\max\{0,re^{-2f(t)}-\frac{1-r}{2}\},\nonumber\\
B^{\Phi}_{\max}(\hat{\rho})&=&=2r\sqrt{1+ e^{-4f(t)}}.
\end{eqnarray}
For the special value of $\alpha=1/\sqrt{2}$, the dynamics of the entanglement and the Bell nonlocality  were investigated by Li and Liang who showed that $C(\hat{\rho})$ and $B_{\max}(\hat{\rho})$ show exactly same dynamics for $\hat{\rho}_{\Phi}(0)$ and $\hat{\rho}_{\Psi}(0)$ initial states in the absence of dipole-dipole interaction~\cite{ll}, thus we will not discuss it here.
\begin{figure}[!ht]\centering
\includegraphics[width=6.5cm]{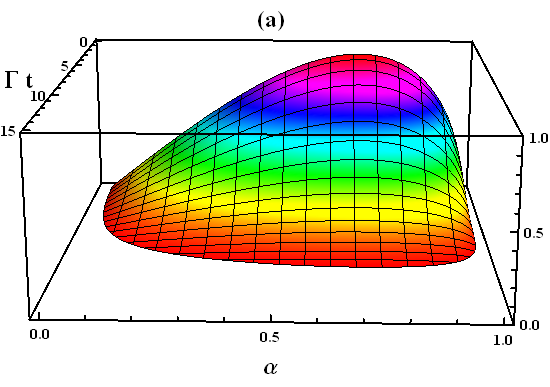}
\includegraphics[width=6.5cm]{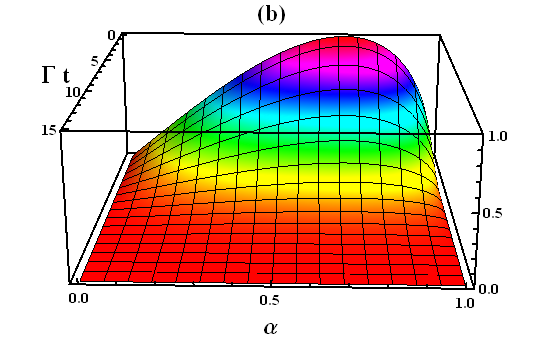}

\includegraphics[width=6.5cm]{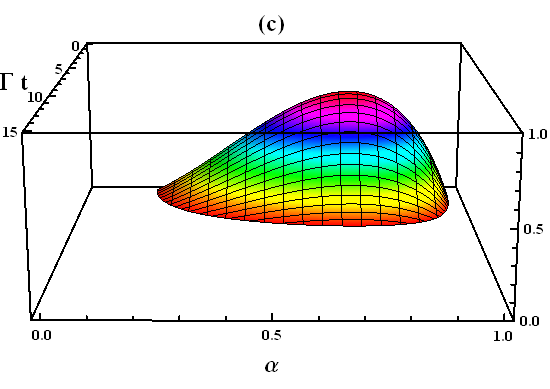}
\includegraphics[width=6.5cm]{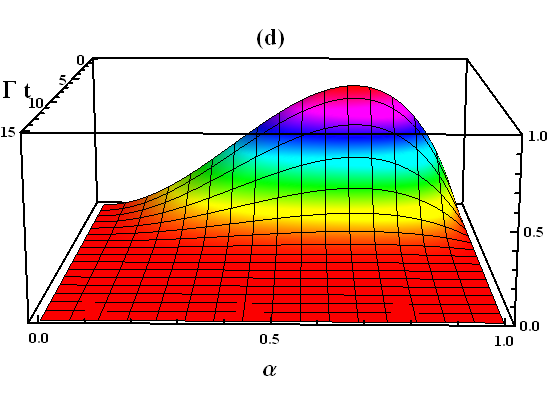}
\caption{\label{fig.4} $C(\hat{\rho})$~(Fig.~(a) and~(b)) and $B_{\max}(\hat{\rho})-2$~(Fig.~(c) and~(d)) versus $\Gamma t$ and $\alpha$  for $\hat{\rho}_{\Phi}(0)$ or $\hat{\rho}_{\Psi}(0)$ with  $J=0,\gamma/\Gamma=0.1$ and $r=0.95$~(Fig.~(a) and~(c)) and $r=1$~(Fig.~(b) and~(d)).}
\end{figure}

To further elucidate the role of the so called  degree of entanglement parameter $\alpha$ in the dynamics of $C(\hat{\rho})$ and $B_{\max}(\hat{\rho})$, we consider the general initial states given by Eq.~(\ref{ewl}). It is easy to write down the time-dependent density matrix elements for the type of initial states of Eq.~(\ref{ewl}) by using the solution in~\ref{App1}. The dynamics of concurrence and Bell nonlocality as a function of the parameter $\alpha$ of the initial state are presented in Fig.~4 and~5. In Fig.~4, we display the behavior of $C(\hat{\rho})$ and $B_{\max}(\hat{\rho})-2$ for the non-interacting case for the initial state $\hat{\rho}_{\Phi}(0)$ or $\hat{\rho}_{\Psi}(0)$ , while Fig.~5 presents the dynamics for the $J=0.5$ value of the dipole-dipole interaction strength for $\hat{\rho}_{\Phi}(0)$. For the non-interacting case, the effect of purity and the $\alpha$ parameter on the dynamics of  $C(\hat{\rho})$ and $B_{\max}(\hat{\rho})-2$ are displayed in Figs.~4(a) and~(b) and Figs.~4(c) and~(d), respectively. As can be seen from these figures, the entanglement and Bell nonlocality are maximum for $\alpha=1/\sqrt{2}$ which corresponds to initial Bell-state which offers the longest time of survival for the mixed state case~(Fig.~4(a) and~(c)) in the absence of dipole-dipole interaction. The volume of nonzero entanglement and Bell nonlocality decreases as $r$~(purity) decreases as can be seen from a comparison of Figs.~4(a) and~4(b) and also~4(c) and~4(d). For $r=1$, the entanglement and Bell nonlocality have non-zero values for all $\alpha$ values with exponential decay, except for $\alpha=0$ or $\alpha=1$.  One should also note that Figs.~4(a)-(d) are for non-Markovian dynamics, in the Markovian case Fig.~4(b) and~(d) would be qualitatively similar, but the nonzero regions in  Fig.~4(a) and~(c) would shrink~\cite{noiseonly,fare,ll}.
\begin{figure}[!ht]\centering
\includegraphics[width=6.5cm]{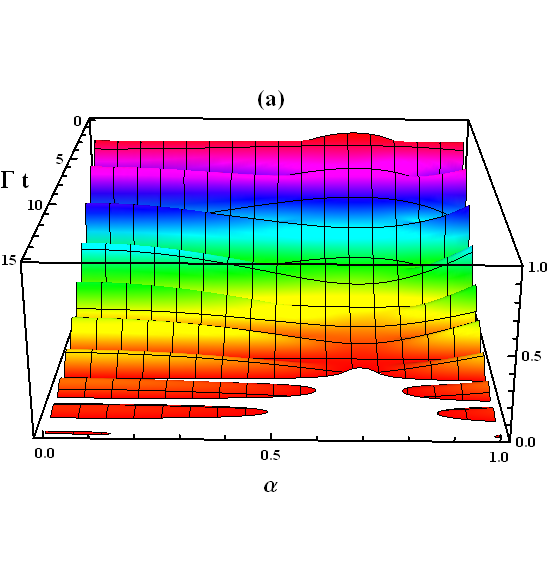}
\includegraphics[width=6.5cm]{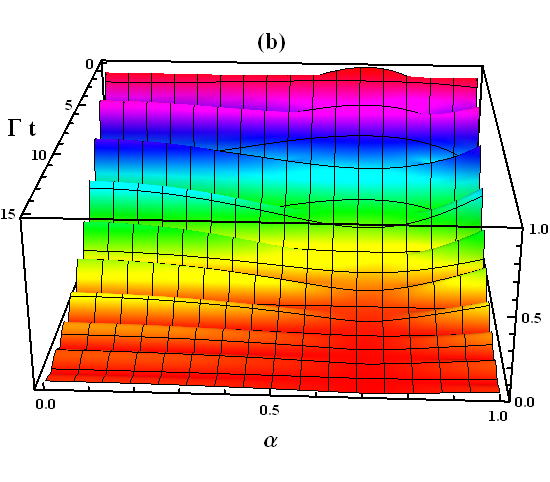}

\includegraphics[width=6.5cm]{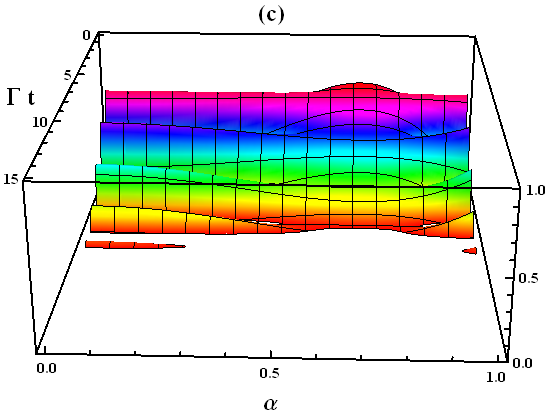}
\includegraphics[width=6.5cm]{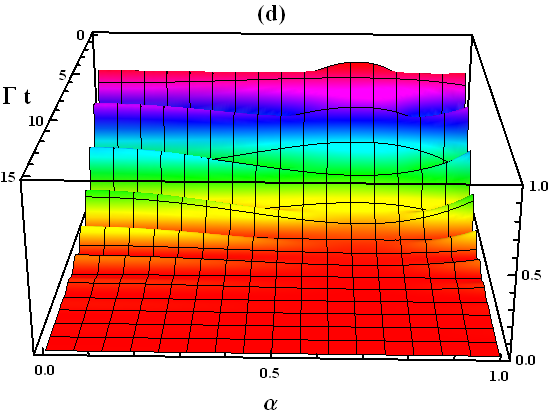}
\caption{\label{fig.5} $C(\hat{\rho})$~(Fig.~(a) and~(b)) and $B_{\max}(\hat{\rho})-2$~(Fig.~(c) and~(d)) versus $\Gamma t$ and $\alpha$  for $\hat{\rho}_{\Phi}(0)$  with  $J=0.5,\gamma/\Gamma=0.1$ and $r=0.95$~(Fig.~(a) and~(c)) and $r=1$~(Fig.~(b) and~(d)).}
\end{figure}

The effect of dipole-dipole interaction on the dynamics  $C(\hat{\rho})$ and $B_{\max}(\hat{\rho})-2$ as function of $\alpha$ and $\Gamma t$ are displayed in Figs.~5(a)-(d) for $\hat{\rho}_{\Phi}(0)$ with  $J=0.5$. Comparing to the corresponding subfigures of Fig.~4 and~5, the most pronounced difference is in the dynamics of the initial state with $\alpha=1/\sqrt{2}$ which suffers the quickest death among all the other probable initial mixed states. Also $\alpha=0$ and $\alpha=1$ states which are initially non-entangled are found to have entanglement for the longest time under dipole-dipole interaction. Also we should mention that for the pure initial states~($r=1$), the dipole-dipole interaction leads to non-zero entanglement and violation of Bell inequality for $\alpha=0$ and $\alpha=1$ for all times as can be seen from Figs.~5(b) and~5(d).
\section{Conclusion}
\label{conc}
We have investigated the dynamics of entanglement as measured by concurrence and Bell nonlocality as measured by CHSH inequality for two noisy qubits which are connected to each other by dipole-dipole interaction. The considered noise have Ornstein-Uhlenbeck type correlation and the dynamics is investigated analytically for extended Werner-like initial states. We have found that the most important effects of dipole-dipole interaction to the dynamics of the considered quantities are as follows: The dipole-dipole interaction can create nonzero entanglement and lead to violation of Bell inequality for $\hat{\rho}_{\Phi}(0)=\frac{1-r}{4}I_4+r\left|\Phi\right\rangle\left\langle \Phi\right|$ initial state~(where $\left|\Phi\right\rangle=\alpha\left|10\right\rangle+\sqrt{1-\alpha^2}\left|01\right\rangle$) with $\alpha=0$ or $\alpha=1$ which can be protected from  sudden death for a  pure initial state~($r=1$). On the other hand, for mixed initial states~($r<1$) the dynamics of $\hat{\rho}_{\Phi}(0)$ with $\alpha=1/\sqrt{2}$ is found to have longest life time for nonzero entanglement and Bell nonlocality violation in the absence of dipole-dipole interaction. The considered quantities suffer the quickest death for $\alpha=1/\sqrt{2}$ if the dipole-dipole interaction between qubits is considered.

\appendix
\section{}
\label{App1}
The solutions of the differential equations~(\ref{maselements}) can be obtained as
\begin{eqnarray}
\label{solmaselements}
\rho_{ii}&=&\rho_{ii}(0)\quad(i=1,4),\nonumber\\
\rho_{22}&=&\frac{Y+Z}{2},\nonumber\\
\rho_{33}&=&\frac{Y-Z}{2},\nonumber\\
\rho_{12}&=&(\rho_{12}(0)\cos(2Jt)+i\rho_{13}(0)\sin(2Jt))e^{-f(t)},\nonumber\\
\rho_{13}&=&(\rho_{13}(0)\cos(2Jt)+i\rho_{12}(0)\sin(2Jt))e^{-f(t)},\nonumber\\
\rho_{14}&=&\rho_{14}(0)e^{-2f(t)},\nonumber\\
\rho_{23}&=&\frac{A+B}{2},\nonumber\\
\rho_{24}&=&(\rho_{24}(0)\cos(2Jt)-i\rho_{34}(0)\sin(2Jt))e^{-f(t)},\nonumber\\
\rho_{34}&=&(\rho_{34}(0)\cos(2Jt)-i\rho_{24}(0)\sin(2Jt))e^{-f(t)},
\end{eqnarray}
where
\begin{eqnarray}
f(t)&=&\int_0^tG(s)ds\nonumber\\
&=&\frac{\Gamma}{2}\left(t+\frac{1}{\gamma}(e^{-\gamma t}-1)\right),\nonumber\\
Y&=&\rho_{22}(0)+\rho_{33}(0),\nonumber\\
Z&=&\frac{ K\left(\frac{e^{-\gamma t}}{J^2}\right)^{\eta_+}{_1F_1}\left(\eta_+;1+\frac{\epsilon}{\gamma};-\frac{\Gamma e^{-\gamma t}}{\gamma}\right)+L\left(\frac{1}{J^2}\right)^\frac{\epsilon}{\gamma}\left(\frac{e^{-\gamma t}}{J^2}\right)^{\eta_-}{_1F_1}\left(\eta_-;1-\frac{\epsilon}{\gamma};-\frac{\Gamma e^{-\gamma t}}{\gamma}\right)}{M},\nonumber\\
A&=&(\rho_{23}(0)+\rho_{23}^*(0))e^{-2f(t)},\nonumber\\
B&=&\left(\frac{1}{J^2}\right)^{-\eta_+} \left(\frac{e^{-\gamma t}}{J^2}\right)^{\eta_-}\left(\frac{C+D}{E}\right),
\end{eqnarray}
where
\begin{eqnarray}
K&=&-\Gamma(\rho_{22}(0)-\rho_{33}(0))(\gamma(\gamma-\Gamma)(\epsilon-\Gamma)-64J^2\gamma){_1F_1}\left(\kappa_-;2-\frac{\epsilon}{\gamma};-\frac{\Gamma}{\gamma}\right)\nonumber\\
&+&\gamma(\gamma^2-\epsilon^2){_1F_1}\left(\eta_-;1-\frac{\epsilon}{\gamma};-\frac{\Gamma}{\gamma}\right)((\rho_{22}(0)-\rho_{33}(0))(\epsilon-\Gamma)\nonumber\\
&-&8iJ(\rho_{23}(0)-\rho_{32}(0))),\nonumber\\
L&=&\Gamma(\rho_{22}(0)-\rho_{33}(0))(\gamma(\Gamma-\gamma)(\epsilon+\Gamma)-64J^2\gamma){_1F_1}\left(\kappa_+;2+\frac{\epsilon}{\gamma};-\frac{\Gamma}{\gamma}\right)\nonumber\\
&+&\gamma(\gamma^2-\epsilon^2){_1F_1}\left(\eta_+;1+\frac{\epsilon}{\gamma};-\frac{\Gamma}{\gamma}\right)((\rho_{22}(0)-\rho_{33}(0))(\epsilon+\Gamma)\nonumber\\
&+&8iJ(\rho_{23}(0)-\rho_{32}(0))),\nonumber\\
M&=&\left(\frac{1}{J^2}\right)^{\eta_+}(_1F_1\left(\eta_-;1-\frac{\epsilon}{\gamma};-\frac{\Gamma}{\gamma}\right)(2\gamma\epsilon(\gamma^2-\epsilon^2){_1F_1}\left(\eta_+;1+\frac{\epsilon}{\gamma};-\frac{\Gamma}{\gamma}\right)\nonumber\\
&+&\Gamma(\gamma(\Gamma-\gamma)(\Gamma+\epsilon)-64J^2\gamma){_1F_1}\left(\kappa_+;2+\frac{\epsilon}{\gamma};-\frac{\Gamma}{\gamma}\right))-\Gamma(\gamma(\gamma-\Gamma)(\epsilon-\Gamma)\nonumber\\
&-&64J^2\gamma){_1F_1}\left(\eta_+;1+\frac{\epsilon}{\gamma};-\frac{\Gamma}{\gamma}\right){_1F_1}\left(\kappa_-;2-\frac{\epsilon}{\gamma};-\frac{\Gamma}{\gamma}\right)),\nonumber\\
C&=&\left(\frac{1}{J^2}\right)^{\frac{\epsilon}{\gamma}}{_1F_1}\left(\kappa_-;1-\frac{\epsilon}{\gamma};-\frac{e^{-\gamma t}\Gamma}{\gamma}\right)(\gamma\Gamma(\rho_{32}(0)-\rho_{23}(0))((\gamma+\Gamma)(\Gamma-2\gamma+\epsilon)\nonumber\\
&-&64J^2){_1F_1}\left(\Delta_+;2+\frac{\epsilon}{\gamma};-\frac{\Gamma}{\gamma}\right)+\gamma(\gamma^2-\epsilon^2){_1F_1}\left(\kappa_+;1+\frac{\epsilon}{\gamma};-\frac{\Gamma}{\gamma}\right)\nonumber\\
&\times&((\rho_{32}(0)-\rho_{23}(0))(\Gamma+\epsilon)-8iJ(\rho_{22}(0)-\rho_{33}(0)))),\nonumber\\
D&=&\left(\frac{e^{-\gamma t}}{J^2}\right)^{\frac{\epsilon}{\gamma}}{_1F_1}\left(\kappa_+;1+\frac{\epsilon}{\gamma};-\frac{e^{-\gamma t}\Gamma}{\gamma}\right)(\gamma\Gamma(\rho_{32}(0)-\rho_{23}(0))((\gamma+\Gamma)(2\gamma-\Gamma+\epsilon)\nonumber\\
&+&64J^2){_1F_1}\left(\Delta_-;2-\frac{\epsilon}{\gamma};-\frac{\Gamma}{\gamma}\right)+\gamma(\gamma^2-\epsilon^2){_1F_1}\left(\kappa_-;1-\frac{\epsilon}{\gamma};-\frac{\Gamma}{\gamma}\right)\nonumber\\
&\times&((\rho_{23}(0)-\rho_{32}(0))(\Gamma-\epsilon)+8iJ(\rho_{22}(0)-\rho_{33}(0)))),\nonumber\\
E&=&{_1F_1}\left(\kappa_-;1-\frac{\epsilon}{\gamma};-\frac{\Gamma}{\gamma}\right)(-2\gamma\epsilon(\gamma^2-\epsilon^2){_1F_1}\left(\kappa_+;1+\frac{\epsilon}{\gamma};-\frac{\Gamma}{\gamma}\right)\nonumber\\
&-&\gamma\Gamma((\gamma+\Gamma)(\Gamma-2\gamma+\epsilon)-64J^2){_1F_1}\left(\Delta_+;2+\frac{\epsilon}{\gamma};-\frac{\Gamma}{\gamma}\right))\nonumber\\
&-&\gamma\Gamma\left((\gamma+\Gamma)(2\gamma-\Gamma+\epsilon)+64J^2\right){_1F_1}\left(\Delta_-;2-\frac{\epsilon}{\gamma};-\frac{\Gamma}{\gamma}\right) {_1F_1}\left(\kappa_+;1+\frac{\epsilon}{\gamma};-\frac{\Gamma}{\gamma}\right),\nonumber\\
\end{eqnarray}
where $\epsilon=\sqrt{\Gamma^2-64J^2},\kappa_\pm=\frac{2\gamma+\Gamma\pm\epsilon}{2\gamma},\Delta_\pm=\frac{4\gamma+\Gamma\pm\epsilon}{2\gamma},\eta_{\pm}=\frac{\Gamma\pm\epsilon}{2\gamma}$ and ${_1F_1}(a;b;z)$ is the Kummer confluent hypergeometric function~\cite{mos}. These solutions are obtained by using Mathematica program.

\end{document}